\documentclass[aps,preprint,tightenlines]{revtex4}
\usepackage{graphicx}

\begin{document}
\preprint{\rm FIU-NUPAR-\today{}\\}

\medskip
\medskip

\title{HARD PHOTODISINTEGRATION OF $^3$He}

\author{CARLOS GRANADOS$^*$}

\affiliation{Florida International University\\
Miami, Florida 33199, U.S.\\
$^*$E-mail: cgran005@fiu.edu\\
www.fiu.edu}

\begin{abstract}
Large angle photodisintegration of two nucleons from the $^3$He nucleus is studied within the framework of the hard rescattering model (HRM). In the HRM the incoming photon is absorbed by one nucleon's valence quark that then undergoes a hard rescattering reaction with a valence quark from the second nucleon producing two nucleons emerging at large transverse momentum . Parameter free cross sections for pp and pn break up channels are calculated through the input of experimental cross sections on $pp$ and $pn$ elastic scattering.  The calculated cross section for $pp$ breakup  and its predicted energy dependency are in good agreement with recent experimental data. Predictions on spectator momentum distributions and helicity transfer are also presented. 
\end{abstract}

\maketitle

\section{Introduction}\label{aba:sec1}
Breakup of nucleon-nucleon ($NN$) systems at large energy and momentum transfer is an important tool in probing QCD descriptions of nuclear processes at this kinematic regime. In $NN$ breakup, large $NN$ center of mass energies where QCD degrees of freedom become evident, are reached with moderate beam energies in comparison with nucleon nucleon collisions.  The simplest of these processes, the photodisintegration of the deuteron has been intensively studied through experiments at kinematics where the hadronic description is no longer applicable (see, e.g., Refs.\cite{Bochna:1998ca},\cite{Schulte:2001se}). It has been observed that at 90$^o$ center of mass angle and beam energies of few GeV, the measured differential cross section scales in accordance with the constituent counting 
rule for exclusive processes \cite{Brodsky:1973kr}, signaling the onset of quark degrees of freedom in the reaction.

 Several QCD models of $NN$ breakup have been developed in order to described features observed in the experimental data. Among these, models such as the reduced nuclear amplitude (RNA) approach \cite{Brodsky:1983kb}, the quark gluon string (QGS) model \cite{Grishina:2001cr}, and the hard rescattering model (HRM) \cite{Frankfurt:1999ik} rely on different approaches in dealing with the nonperturbative factors that enter the calculations. For the RNA approach, these factors are accounted for by introducing parameterizations of nucleon and nuclear form factors. The QGS model introduces a reggeization of the scattering amplitude, while through the HRM, breakup cross sections are calculated through the input of experimental data of cross sections of the corresponding nucleon nucleon elastic scattering process.
 
Calculations of cross sections of deuteron photodisintegration in the RNA and the QGS approaches require a normalization to the data. In contrast, in the HRM these cross sections are obtained without the need of a normalization parameter. In order to further test the validity of these competing approaches, it was proposed to extend photodisintegration studies to the breakup of proton proton ($pp$) and proton neutron ($pn$) systems in the $^3$He nucleus \cite{Brodsky:2003ip}. For these reactions, numerical predictions for the cross section differ considerably from one approach to another. Recent experimental data \cite{Pomerantz:2009sb} shows a better agreement with HRM calculations\cite{Sargsian:2008zm}.

Additional constrains on the analysis of $NN$ breakup are set in $^3He$ photodisintegration by studying observables such as the spectator momentum distributions and polarization transfer asymmetries. Features on these observables predicted here within the HRM are expected to be put to the test in future experimental analysis.

.

\section{HRM of $^3He$ Photodidisintegration}
\begin{figure}[ht]
\centering\includegraphics[height=3cm,width=4.5cm]{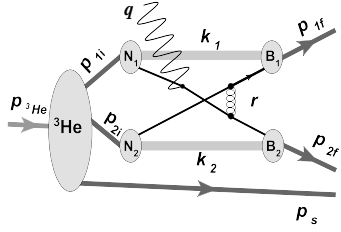}
\caption{$^3$He photodisintegration according to the HRM}
\label{fig:hrmf}
\end{figure}

Fig.\ref{fig:hrmf} illustrates the HRM picture of the breakup of a $NN$ pair in $^3$He photodisintegration. It differs  graphically from that of the deuteron breakup by the presence of a spectator nucleon. In Fig.\ref{fig:hrmf}, by absorbing  the incoming photon, a valence quark  is knocked out from one of the nucleons in the initial $NN$ system. This quark then re scatters off a valence quark from the second nucleon. Both quarks then recombine with the spectators systems to form two nucleons with large transverse momentum in the final state. 

Within the HRM, The scattering amplitude  is obtained  using Feynman rules on the diagram in Fig.\ref{fig:hrmf}. The only calculable terms are the QED vertex corresponding to the photon absorption by the valence quark and the component of the nuclear wave function represented by the $^3$He vertex. The scattering amplitude is then reduced to the form in Eq.(\ref{ampl}) through two main assumptions (see Ref. \cite{Sargsian:2008zm} for details). First, by working in a $q^+=0$ reference frame, we approximate the struck nucleon's propagator to its on shell component. Second, we assume that the nucleons in the $NN$ system have an equal share of the $NN$ light cone momentum, which maximizes the nuclear wave function. The result as shown in Eq.(\ref{ampl}) is a convolution of a nuclear wave function $\Psi_{^3\textnormal{\scriptsize {He}}}$ and the scattering amplitude $T$ of a kinematically corresponding $NN$ elastic scattering reaction.  
\begin{eqnarray}
\langle\lambda_{1f},\lambda_{2f},\lambda_s\mid {\cal M} \mid \lambda_\gamma, \lambda_A\rangle 
 =&  ie[\lambda_\gamma]\times \nonumber\\
 & \left\{ \sum\limits_{i \in N_1}\sum\limits_{\lambda_{2i}} \int 
{Q_i^{N_1}\over \sqrt{2s'}}
\langle \lambda_{2f};\lambda_{1f}\mid T^{QI}_{NN,i}(s,t_N)\mid 
\lambda_\gamma;\lambda_{2i}\rangle\right.   \nonumber \\
&\left.\Psi_{^3\textnormal{\scriptsize {He}}}^{\lambda_A}(p_1,\lambda_\gamma;p_2,\lambda_{2i};p_s,\lambda_s)
{d^2p_\perp \over (2\pi)^2} \right.  \nonumber \\ 
&  + \left.
\sum\limits_{i \in N_2}\sum\limits_{\lambda_{1i}} \int 
{Q_i^{N_2}\over \sqrt{2s'}}
\langle \lambda_{2f};\lambda_{1f}\mid T^{QI}_{NN,i}(s, t_{N})\mid \lambda_{1i};
\lambda_{\gamma}\rangle \right.  \nonumber\\
&\left.\Psi_{^3\textnormal{\scriptsize {He}}}^{\lambda_A}(p_1,\lambda_{1i};p_2,\lambda_{\gamma};p_s,\lambda_s)
{d^2p_\perp \over (2\pi)^2} \right\}, \nonumber\\
\label{ampl}
\end{eqnarray}
where $\lambda_1,\lambda_2,\lambda_s$ are the helicities of the nucleons from $^3$He. $\lambda_A,\lambda_\gamma$ are the helicities of the nucleus and of the incoming photon respectively, and $Q^{N}_i$ is the charge in $e$ units of the quark struck by the photon. 
A further reduction is achieved if $-t_N>>m_N^2$. Under this condition, the matrix elements of $T$ can be factored out of the $p_\perp$ integral.
   
Using the following notation for the independent helicity amplitudes in $NN$ elastic scattering,
\begin{eqnarray}
\left\langle +,+ \left|T\right|+,+\right\rangle &=& \phi_1\nonumber\\
\left\langle +,- \left|T\right|+,-\right\rangle &=& \phi_3\nonumber\\
\left\langle -,+ \left|T\right|+,-\right\rangle &=& \phi_4\nonumber\\
\left\langle -,- \left|T\right|+,+\right\rangle &=& \phi_2\nonumber\\
\left\langle -,+ \left|T\right|+,+\right\rangle &=& \phi_5,\nonumber\\
\label{pnham}
\end{eqnarray}
and from explicitly working the matrix elements in Eq.(\ref{ampl}), we can write the averaged  of the squared of the scattering amplitudes ${\cal M}$ as follows,

\begin{eqnarray}
\bar{|{\cal M}|^2}&=&{1\over2}{1\over3}{e^2\over 2s^\prime}\left[S_{12}\left\{|(\hat{Q}^{N_1}+\hat{Q}^{N_2})\phi_1|^2+
|(\hat{Q}^{N_1}+\hat{Q}^{N_2})\phi_2|^2\right\}\right.\nonumber\\
&&+\left.S_{34}\left\{|\hat{Q}^{N_1}\phi_3+\hat{Q}^{N_2}\phi_4|^2+|\hat{Q}^{N_1}\phi_4+
\hat{Q}^{N_2}\phi_3|^2\right\}\right.\nonumber\\
&&+\left.2S_0|(\hat{Q}^{N_1}+\hat{Q}^{N_2})\phi_5|^2\right],
\label{asamp}
\end{eqnarray}
where the light-cone spectral functions of the $^3$He nucleus  are defined as follows:
\begin{eqnarray}
S_{12}&=&N_{NN}\sum^{1\over2}_{(\lambda_1=\lambda_2=-{1\over2})}\sum^{1\over2}_{(\lambda_s=-{1\over2})}
\left|\int\Psi_{^3\textnormal{\scriptsize {He}}}^{1\over2}(p_1,\lambda_1;p_2,\lambda_2;p_s,\lambda_s)
{d^2p_\perp \over (2\pi)^2}\right|^2,\nonumber\\
S_{34}&=&N_{NN}\sum^{1\over2}_{(\lambda_1=-\lambda_2=-{1\over2})}\sum^{1\over2}_{(\lambda_s=-{1\over2})}\left
|\int\Psi_{^3\textnormal{\scriptsize {He}}}^{1\over2}(p_1,\lambda_1;p_2,\lambda_2;p_s,\lambda_s)
{d^2p_\perp \over (2\pi)^2}\right|^2,\nonumber\\
S_0&=&S_{12}+S_{34},
\label{spfun}
\end{eqnarray}
and the charge factors ${\hat  Q}^N$ account for all possible quark photon interactions contributing to the reaction's amplitude.
For two nucleon breakup in $^3$He, the unpolarized cross section is given by
\begin{equation}
{d\sigma\over dt d^3p_{s}/(2E_s (2\pi)^3)} = {|\bar {\cal M}|^2\over 16\pi (s-M_A^2)(s_{NN}-M^2_{NN})}
\label{hrm_crs}
\end{equation}
where $s = (q + p_{A})^2$, and $s_{NN}=(q+p_{NN})^2$. Then according to Eq.(\ref{asamp}), the cross section in Eq.(\ref{hrm_crs}) can be computed   from the knowledge of the $\phi$ scattering amplitudes, and of the charge factors ${\hat Q}$. Because there is not enough experimental data in polarized $NN$ scattering to extract the helicity amplitudes $\phi$, Eq.(\ref{hrm_crs}) can instead be computed through general assumptions regarding these amplitudes . This is further discussed in the coming sections. 

Alternatively, the $\phi$ amplitudes can be modeled according to a $NN$ scattering mechanism. Experiments have demonstrated the dominance of the quark interchange mechanism (QIM) in hard exclusive reactions between baryons of common quark flavors such as $pp$ and $pn$ elastic scattering \cite{White:1994tj}. Within a (QIM) approach, the $\phi$ amplitudes are obtained by assuming a symmetry structure of the valence quark wave functions of the interacting baryons (see e.g., Refs.\cite{Farrar:1978by}). A starting point is using SU(6) symmetric wave functions \cite{Farrar:1978by}, but it has been suggested that a diquark model better describes general features in $pn$ elastic scattering such as its  strength  relative to $pp$ scattering, and the asymmetry observed in its angular distributions (see, e.g., Ref.\cite{Granados:2009jh}).

Finally, an important feature  of the HRM is that at large values of $s$ and $-t$  it predicts an energy distribution of Eq.(\ref{hrm_crs}) in accordance with the constituent counting rule \cite{Brodsky:1973kr}. That is ${d\sigma\over dt}\sim s^{-11}$ \cite{Frankfurt:1999ik},\cite{Brodsky:2003ip},\cite{Sargsian:2008zm}, which comes from the $s^{-4}$ dependence of the $\phi$ amplitudes entering Eq.(\ref{asamp}). This behavior has already been observed in deuteron photodisintegration \cite{Bochna:1998ca},\cite{Schulte:2001se}, and in recent experiments on $pp$ breakup in $^3$He \cite{Pomerantz:2009sb}.

\subsection{$pn$ and $pp$ breakup}  
As mentioned in the previous section ${d\sigma\over dt}$ in Eq.(\ref{hrm_crs}) can be computed through general assumptions on $pn$ and $pp$ elastic scattering.
\subsubsection{$\gamma ^3$He$\rightarrow pn+p$}
In $pn$ breakup, $pn$ scattering amplitudes $\phi$ enter Eq.(\ref{asamp}). Using  that for hard $pn$ elastic scattering $\phi_3\sim\phi_4$ (see, e.g., Ref. \cite{Ramsey:1991bd}), and that at low transverse momentum $p_\perp<<m_N$, $\Psi_{^3\textnormal{\scriptsize {He}}}\approx\sqrt2(2\pi)^3m_N\Psi_{^3\textnormal{\scriptsize {He,NR}}}$ \cite{Frankfurt:1988nt}, the cross section of $pn$ breakup can be approximated to a simple product of a spectral function and a cross section of a kinematically corresponding $pn$ elastic reaction,

\begin{eqnarray}
{d\sigma^{\gamma ^3He\rightarrow (pn) p}\over dt {d^3p_s\over{E_s}}} =&&  
{\alpha Q^2_{F,pn} 16\pi^4}{S^{pn}_{0,NR}(\alpha_c={1\over 2},\vec p_s)\over 2}
{s_{NN}(s_{NN}-4m_N^2)\over (s')(s-M^2_{^3\textnormal{\scriptsize {He}}})}\nonumber\\
&&\times{d\sigma^{pn\rightarrow pn}(s_{NN},t_N)\over dt_N },
\nonumber \\
\label{pn_hrm_crs}
\end{eqnarray} 
where $Q_F=1/3$, $\alpha=1/137$ and $s'=s_{NN}-M_{NN}^2$.

\subsubsection{$\gamma ^3$He$\rightarrow pp+n$}
\begin{figure}
\begin{minipage}[b]{0.4\linewidth}
\centering
\includegraphics[height=4cm,width=5cm]{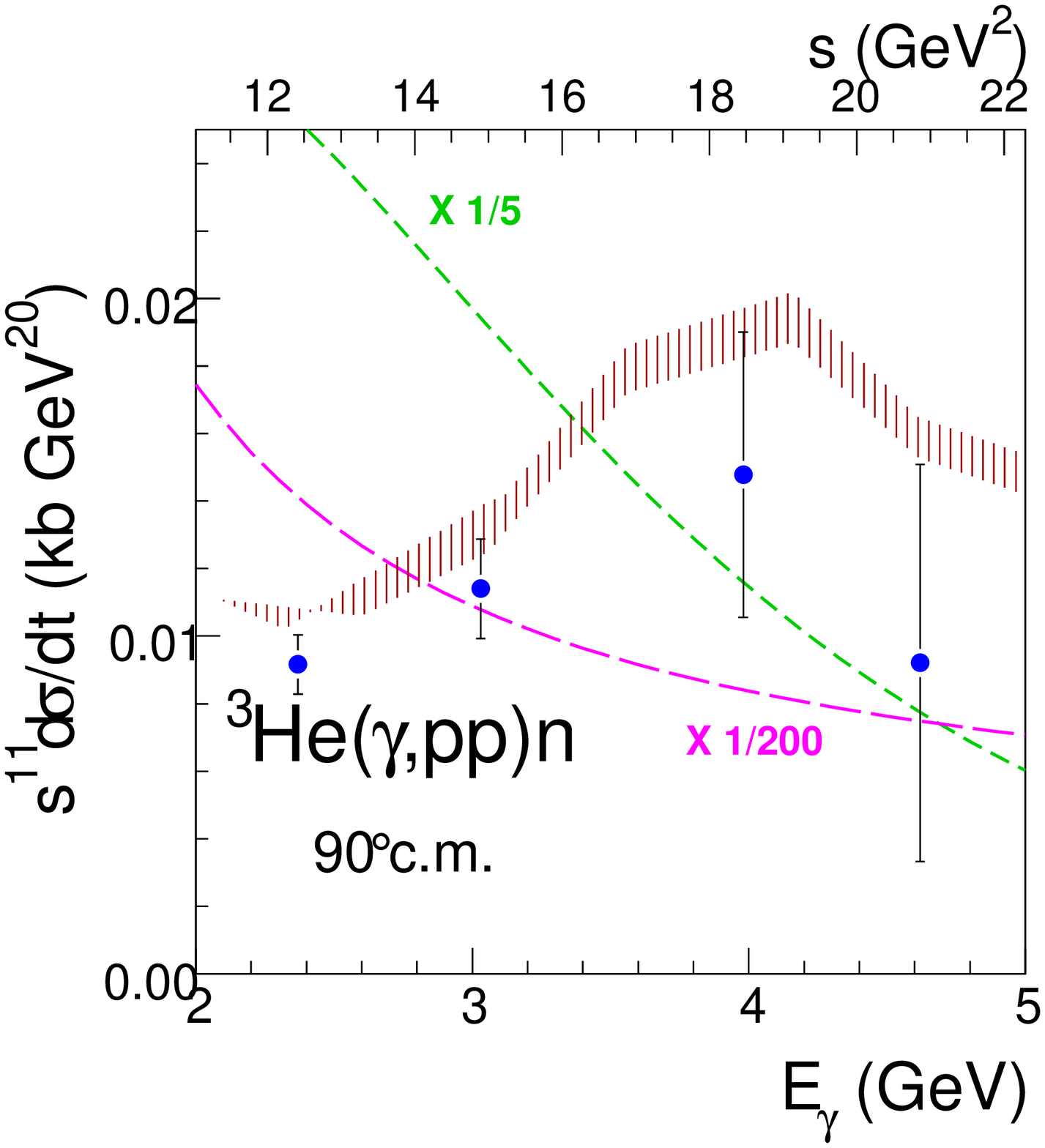}
      \caption{Cross section data for $pp$ breakup in $^3$He photodisintegration at 90$^o$ c.m. angle of the $\gamma+pp$ system \cite{Pomerantz:2009sb}. (Dashed) QGS prediction times 1/5. (Long dashed) RNA prediction times 1/200. (Shaded) HRM prediction.}
\label{fig:gpp}
\end{minipage}
\hspace{0.7cm}
\begin{minipage}[b]{0.3\linewidth}
\centering
\includegraphics[height=3cm,width=4.5cm]{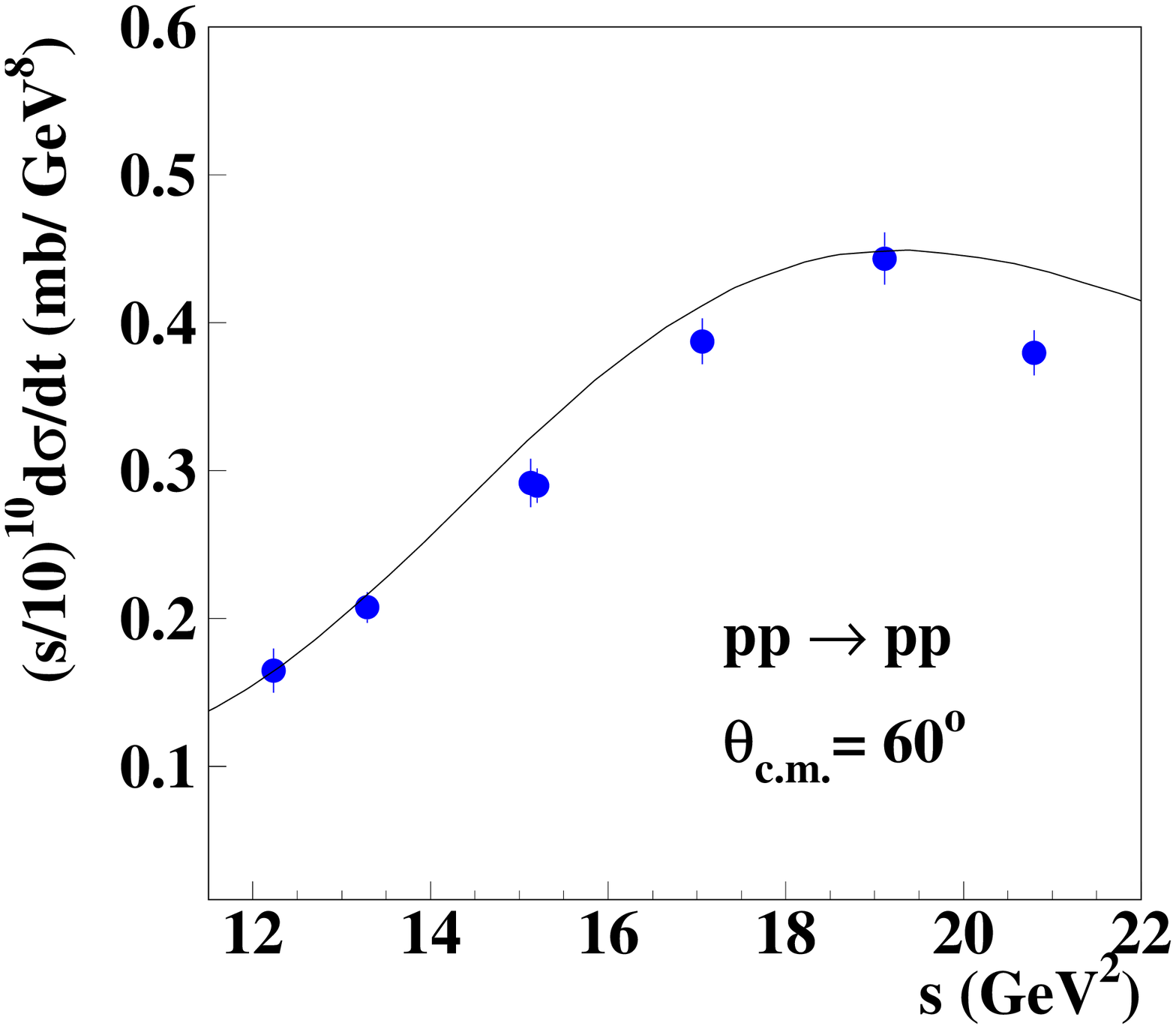}
           \caption{$pp$ elastic scattering data at 60$^o$ of center of mass angle of scattering. These data is used in Eq.(\ref{pp_hrm_crs}) to calculate the HRM cross section of $pp$ breakup shown in Fig.\ref{fig:gpp}.}
      \label{fig:elpp}
      \end{minipage}
\end{figure}      

At large center of mass angle of scattering in $pp$ elastic scattering, it follows from Pauli's principle that $\phi_4\sim-\phi_3$ (see, e.g., Ref.\cite{Jacob:1959at}). Also, assuming that the $S$ states dominates the  ground state nuclear wave function, from the exclusion principle, the probability of two protons having the same helicity is negligible. Therefore, $S_{12}<<S_{34}$. These assumptions allow to write the cross section for $pp$ breakup in $^3$He photodisintegration as follows:
 
\begin{eqnarray}
{d\sigma^{\gamma^3He\rightarrow(pp)n}\over dt {d^3p_{s}\over E_s}} = & &   
\alpha Q_{F,pp}^2 16\pi^4 S^{pp}_{34}(\alpha={1\over 2},\vec p_s)
{s_{NN}(s_{NN}-4m_N^2)\over s' (s-M_{^3\textnormal{\scriptsize {He}}}^2)} \nonumber \\
& &\times{2\beta^2\over 1+2C^2}{ d\sigma^{pp\rightarrow pp}(s_{NN},t_{N})\over dt},
\label{pp_hrm_crs}
\end{eqnarray}
where,
$
C^2=(\phi_3^2/ \phi_1^2)\approx(\phi_4^2/ \phi_1^2)
$,
$
\beta  = (|\phi_3| - |\phi_4|)/|\phi_1|
$ and $Q_F$=5/3. Because  the $pp$ breakup cross section is proportional to $\beta$ and having that $|\phi_4|\sim|\phi_3|$, we predict a large suppression of the $pp$ breakup relative to the $pn$ breakup case.

Eq.(\ref{pp_hrm_crs}) shows that just as in the $pn$ breakup case, the $pp$ breakup cross section is proportional to the cross section of the corresponding  $pp$ elastic scattering. This is illustrated in Figs. \ref{fig:gpp} and \ref{fig:elpp} that show that the shape of the scaled energy distribution for $pp$ breakup mimics that of the scaled energy distribution for  the corresponding $pp$ elastic scattering. Fig.\ref{fig:gpp} also shows that the HRM calculation is in very good agreement with the experimental data.

\subsection{Spectator momentum distributions}
\begin{figure}[ht]
\centering\includegraphics[height=6cm,width=6cm]{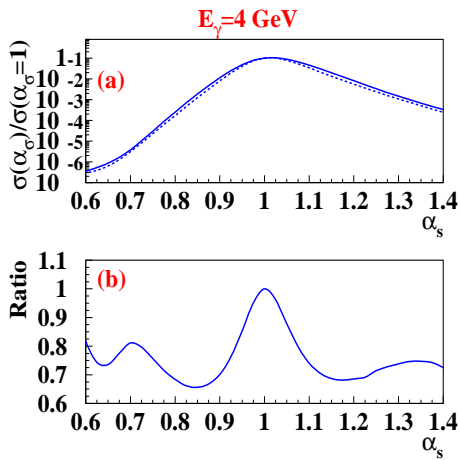}
\caption{(a) $pp$ breakup (solid line) and $pn$ breakup (dashed line)   $s^{11}$-weighted  HRM cross sections as functions of the spectator nucleon's light cone momentum fraction ($\alpha_s$) and normalized to their values at $\alpha_s=1$. (b) Ratio (R) of the $pn$ to $pp$ breakup HRM cross sections  normalized to its value at $\alpha_s=1$. }
\label{fig:ps_dist}
\end{figure}
Additional observables can be extracted from $^3He$ photodisintegration in order to test or constrain the validity of the HRM. We consider for instance the spectator momentum distributions plotted in Fig.\ref{fig:ps_dist}. In Fig.\ref{fig:ps_dist}a we noted that the asymmetry in both distributions around $\alpha_s=1$ is due to the $s^{-11}$ dependence of both cross sections. As less energy is taken by the spectator nucleon, more energy is taken by the $NN$ system, hence the sharper drop towards $\alpha_s=0$. The broader $pp$ breakup distribution  in Fig.\ref{fig:ps_dist}a, as well as the sharp drop of $R$ around $\alpha_s=1$ are due to the suppression of the $S^{pp}_{12}$ spectral function in the low $NN$ momentum component of the nuclear wave function.       
\subsection{Polarization transfer observables}
Another observable of interest is related to the effectiveness with which the incoming photon transfers its helicity to an outgoing  proton. This is measured through the asymmetry $C_{z'}$ (see, e.g., \cite{Sargsian:2008zm}) that in the HRM takes the form
\begin{equation}
C_{z^\prime} = {(|\phi_1|^2 - |\phi_2|^2)S^{12}+ (|\phi_3|^2 - |\phi_4|^2)S_{34}\over
2|\phi_5|^2S_0 + (|\phi_1|^2 + |\phi_2|^2)S_{12}+ (|\phi_3|^2 + |\phi_4|^2)S_{34}},
\label{Cz2}
\end{equation}
Noting that for $pp$ breakup $S_{12}\approx0$ and $\phi_4\approx-\phi_3$, we have that $C^{pp}_{z'}\approx0$, while for $pn$ breakup $C^{pn}_{z'}\approx2/3$.

\section{Summary}
We studied the hard breakup of a nucleon pair in hard a $^3$He nucleus by introducing quark degrees of freedom through the hard rescattering model (HRM). Cross sections for proton neutron $pn$ and proton proton breakup are calculated in the (HRM) without the need for introducing a parameter to normalize to data.
The obtained cross section for $pp$ breakup shows good agreement with recent experimental data.
We also predict distinctive spectator momentum distributions for both $pp$ and $pn$ breakup, and a cancellation of the $C_{z'}$ asymmetry in the $pp$ breakup channel.  Mounting evidence in favor of the rescattering picture of breakup processes motivates an extension of these study to different baryonic channels such as $\Delta\Delta$-isobar in deuteron  breakup (see, e.g., Ref.\cite{Granados:2010cj}) for which experiments are already being proposed.

\bibliographystyle{h-physrev}
\bibliography{myrefs}

\end{document}